\documentclass[10pt, conference, compsocconf]{IEEEtran}

\usepackage{amsmath}
\usepackage{listings } \usepackage{url}
\usepackage{graphicx}

\begin{document}

\title{Collaborative Real Time Coding or How to Avoid the Dreaded Merge}

\author{\IEEEauthorblockN{Stanislav Levin}
\IEEEauthorblockA{The Blavatnik School of Computer Science\\
 Tel Aviv University\\
Tel-Aviv, Israel\\
stanisl@post.tau.ac.il}
\and
\IEEEauthorblockN{Amiram Yehudai}
\IEEEauthorblockA{The Blavatnik School of Computer Science\\
 Tel Aviv University\\
Tel-Aviv, Israel\\
amiramy@tau.ac.il}
}

\maketitle

\begin{abstract}

Software engineers who collaborate to develop software in teams often have to
manually merge changes they made to a module (e.g. a class), because the change
conflicts with one that has just been made by another engineer to the same or
another module (e.g. a supplier class). This is due to the fact that engineers
edit code separately, and loosely coordinate their work via a source control or
a  software configuration management system (SCM).

This work proposes to eliminate almost all the need to manually
merge a recent change, by proposing a Collaborative Real Time Coding
approach. In this approach, valid changes to the code are seen by
others in real time, but intermediate changes (that cause the code
not to compile) result in blocking other engineers from making
changes related to the entity (e.g. method) being modified, while
allowing them to work on most of the system.

The subject of collaborative real time editing systems has been studied for the
past 20 years. Research in this field has mostly concentrated on collaborative
textual and graphical editing. In this work we address the challenges involved
in designing a collaborative real time coding system, as well as present the
major differences when compared to collaborative editing of plain text. We then
present a prototype plug in for the Eclipse Integrated Development Environment (
IDE) that allows for a collaborative coding to take place.

\end{abstract}

\begin{IEEEkeywords}
Collaborative Real Time Coding;  Integrated Development Environment; Software Configuration Management Tools;

\textit{This paper was written in 2011}.
\end{IEEEkeywords}

\section{Introduction}

A major tool used by software engineers to develop software is an
IDE, which appears as a single application that facilitates much of
the engineer's activity by supporting the composition of code,
compiling it, and running tests. Nowadays software engineers usually
work in teams that develop a single software project. This
cooperation is usually accomplished by using a source control
system, or Software Configuration Managementsystem (SCM)(\cite{SVN,
ClearCase, GIT, Mercurial}) . The SCM system maintains all the files
that comprise the software project (here we refer only to program
source files, but the SCM system can also include documentation). An
important task of a SCM is to coordinate between several engineers
who want to modify a file at the same time. While pessimistic
version control models disallow concurrent editing by exclusively
locking files, optimistic version control models leave it up to the
developers to synchronize their actions as to prevent conflicting
editing operations \cite{SurveyOnSoftwareMerging}. In either case,
once the change is done, the engineer commits the file back to the
SCM system, allowing others to check it out and get the most up to
date version of the given file.

We can view the team as sharing a single SCM system, but each having an
individual IDE. While this works quite well, it has some drawbacks. The
different files comprising a software project are interdependent: the code in
one file may refer to entities in other files (in object oriented programming,
files would typically be classes, and they will include calls to methods in
other classes). In the context of object oriented programming it is quite common
that a change in file A may require cascading changes in additional files
\cite{FineGrainedVersinControl}. While an engineer EA is modifying a file A,
engineer EB may be working on file B which has references to A, and may find out
only later, when the modifications of A are committed, that he needs to redo
some of his work to account for the changes in A. The process of merging such
conflicting versions may significantly hinder development, and is extremely
error prone \cite{SurveyOnSoftwareMerging}.  An alternative would be to force EB
to wait with his work on B as long as A is checked out, but this would slow down
the process even if the changes EB wants to do in B do not include any parts
that refer to A.

Manual merges are considered both time consuming and error prone
\cite{ConcurrencyControlInGroupwareSystems}. Since the conflict
involves changes made by multiple users, in order to make the
correct decision a comprehensive understanding of the overall
changes must be obtained. The process of obtaining the information
pertaining to each change may be done in various manners. For
instance, one can query fellow developers about the changes they
made; if the environment supports a change log, it can be inspected
for the change history; some systems may even provide inherent
support, such as the multi versioning technique described in
\cite{MultiVersionConflictResolution}. Regardless of the method
chosen, one thing remains painfully certain - a mishandled merge may
lead to a variety of negative results, ranging from unbuildable code
to a noticeable faulty program behavior, or even worse, an
unnoticeable faulty program behavior. It is not surprising then,
that developers seek to avoid manual merges whenever possible. Once
a conflict is introduced into the system it might be a fairly
complex task to apply an automatic conflict resolution mechanism,
since in many cases several options may seem appropriate from a
syntactic point of view. Semantic based conflict resolution may be
even more complex. It should be noted that the term \textit{merge}
is used both for the task of merging conflicting and non conflicting
code versions. This observation plays an important role, since
although both cases deal with merging several version into one, the
challenges involved in each of them are of different nature.
Recently, distributed SCM tools (\cite{GIT}, \cite{Mercurial}) have
suggested a novel approach to efficiently and automatically merge
non conflicting versions \cite{MovingToDSCM} and thus alleviate the
task of merging. Merging conflicting versions on the other hand,
presents a problem that is unresolvable by automatic means. Merging
conflicting versions is a problem that extends beyond technical
difficulties as it involves merging two changes of which there is no
right and wrong, it just so happened that several changes had taken
place simultaneously and affected the same element. Each change is
syntactically and semantically valid, but only one can be included
in the final version.

Our work addresses the concept of collaborative real time coding by means of
enhancing the existing IDE perception with collaborative real time coding
capabilities. Using our method we hope to minimize the conflicts originating in
poor synchronization between developers, conflicts that otherwise will usually
lead to a \textit{manual} merge process.

The rest of the paper is organized as follows. In Section
\ref{prvwork} we present some related work. We then present our
approach in Section \ref{CRTC}, which includes an illustrative use
case. In Section \ref{prototype} we present our prototype. A
discussion and presentation of future work concludes the paper in
Section \ref{conc}.

\section{Related work}\label{prvwork}

We are not aware of any work done in the field of collaborative real
time coding systems. The subject of collaborative text editing
systems, however, has been studied for the past 20 years and has
mostly concentrated on collaborative textual and graphical editing.
In this section we describe some of the papers that laid the
foundations for collaborative text editing, and inspired us to
further explore the collaborative editing field.

Ellis and Gibbs \cite{ConcurrencyControlInGroupwareSystems}, were the first to
suggest the operational transformation (OT) framework for concurrency management
in a distributed groupware systems. The suggested framework addressed the
difficulties entailed in having a real time, highly reactive, concurrent editing
environment for plain text. The basic idea of OT is to transform arriving
operations against independent \cite{AchievingCCI} operations from the log
(where all previously executed operations are saved) in such a manner, that the
execution of the same set of properly transformed independent operations in
different orders produce identical document states, ensuring convergence
\cite{IssuesInOperationalTransformationInRealTimeGroupEditors}. A major issue
with the correctness of the algorithm presented was what is now commonly
referenced as the dOPT puzzle
\cite{IssuesInOperationalTransformationInRealTimeGroupEditors}. The dOPT puzzle
scenario describes a situation where clients would diverge by more than one step
in their state, breaking the correctness of the algorithm.

In \cite{ConsistencyModel} the consistency model was extended with a
third property: intention violation. The new property stated that if
two operations were independent (as defined by Sun et al
\cite{ConsistencyModel}), then their execution, in any order, must
preserve the intention of each other.

Unlike most other collaborative editing systems, Jupiter
\cite{JupiterCollaborationSystem} implemented a centralized architecture. A
central sever was responsible to mediate between any two clients, so that at any
given time only 2-way communication could take place (the central server and
some client). This server also held the responsibility of propagating changes to
all other clients. When the cental server received an operation request, it was
transformed, if necessary, and applied to the local document state, followed by
propagation to other clients. This centralized manner of communication
inherently relieved Jupiter of both the dOPT puzzle and precedence issues. The
fact that at any given time only a 2-way synchronization took place alleviated
the issue of preserving precedence between operations.

\section{Designing a Collaborative Real Time Coding System\label{CRTC}}

\subsection{Conflicts in Collaborative Work}

We call the state in which the code fails to compile a \emph{buildbreaking}
state or \emph{unbuildable} state, and the state in which the code successfully
compiles a \emph{buildable} state. Similarly, a \emph{buildbreaking change} is a
change that upon execution renders a buildable state unbuildable.

Sun and Sosi\v{c} \cite{OptimalLockingIntegratedWithOperationalTransformation},
describe two different inconsistency categories that may arise in a
collaborative editing system. One is of a generic nature, while the other is
context specific. The generic one deals with inconsistency issues involved in
maintaining the so called {\em CCI} model properties: Convergence, Causality
violation, and Intention preservation \cite{AchievingCCI}. These properties
should be preserved in order to assure correctness as detailed in
\cite{AchievingCCI}. The context specific inconsistency issues stem from the
fact that a particular application domain may have its own, domain specific
rules, as to the validity of its content. In text, for instance, such rules may
include grammatical rules, spelling, etc. In our case, where the application
domain is programming languages, the validity of the content is determined by
whether it is buildable (i.e., whether it successfully compiles). The
operational transformation framework per se deals only with generic consistency,
leaving out context specific challenges. Locking schemes (as detailed in section
\ref{lockingSchemes}) on the other hand are well suited for handling context
specific inconsistencies by providing means of access control and restrictions
on the objects and operations that can be performed in a concurrent fashion.
Proactive approaches aim at preventing a conflict before it actually occurs,
trying to avoid arriving at the conflicted state altogether. While locking
elements for editing, for instance, may provide a means for handling conflicting
operations, many collaborative real time text editing systems opt for a reactive
approach. The reactive approach usually aims at assisting users to resolve a
conflict after it has already occurred, rather than preventing it in the first
place. Once a conflict is detected, the system may provide users with detailed
information as to the conflicting operations, possible resulting states and so
on, leaving it up to the users to resolve the conflict and ultimately decide
upon the desired final document state. Such a reactive approach was employed,
for example, in \cite{MultiVersionConflictResolution} which suggested the
Multi-Versioning technique where several simultaneous versions of the same
object are kept in case of conflicting operations.

\subsection{A Use Case}

We shall now examine a use case demonstrating the CRTC in a real life scenario.

\begin{enumerate}

\item Two developers, Bob and John begin in the same file state.
See figure \ref{fig:UseCaseStep1}.

\begin{figure*}
\center
\caption{Bob and John begin in the same state.}
\label{fig:UseCaseStep1}
\includegraphics[scale=1]{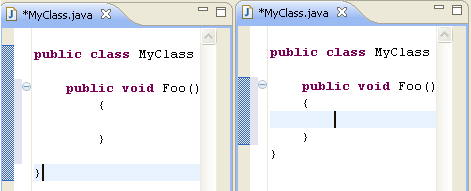}
\end{figure*}

\item Bob intends to add a new parameter, "newParam", of type int to
method "Foo". He begins typing in his change, but mistakenly types in "in
newParam", missing the "t" in the type "int".

\item Not being aware of his
mistyping, Bob also changes the name of the method from "Foo" to "Foo1".
At this stage developer1 has a new name for the method "Foo" (i.e. "Foo1") and
an additional, incomplete parameter definition that renders the file state
unbuildable. Since the file state is currently unbuildable, none of the changes
Bob has made are propagated to the rest of the developers.
See figure \ref{fig:UseCaseStep2}.

\begin{figure*}
\center
\caption{Bob introduces some changes. Method's name is changed, and the
new parameter has an invalid type.}
\label{fig:UseCaseStep2}
\includegraphics[scale=1]{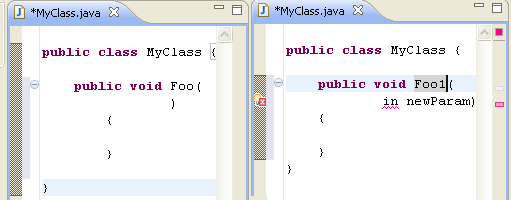}
\end{figure*}

\item
Meanwhile, John intends to change method "Foo" to "Foo2". He is currently
unaware that developer1 has already changed this method's name to "Foo1", and in
the file version he currently has, the method's name is still "Foo". John
begins changing the name of "Foo" to "Foo2", say by typing an additional "2" at
the end of the "Foo" string in the methods definition, and is immediately warned
that the current method is locked for editing by another developer.

 \item
John is now \textit{made aware} that the method is undergoing changes by
some other developer, and may choose to wait till these changes are complete,
avoiding the conflict that would have otherwise been introduced by a concurrent
method name changing by the two developers.
See figure \ref{fig:UseCaseStep3}.

\begin{figure*}
\center
\caption{John tries to change the method's name and is notified that it is
already being changed by another developer.}
\label{fig:UseCaseStep3}
\includegraphics[scale=1]{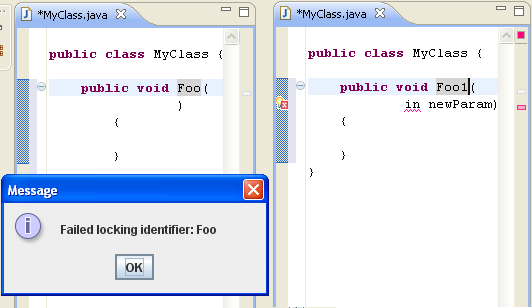}
\end{figure*}

 \item Once Bob adds the "t"
to the mistyped "in", his file becomes buildable, and is instantaneously propagated to all
developers, including developer2, which gets the new method name, with the
additional parameter added by Bob. John may now commence the change
he's been intending.
See figure \ref{fig:UseCaseStep4}.

\begin{figure*}
\center
\caption{Bob fixes the errors and turns file state buildable, the code is
then propagated to John.}
\label{fig:UseCaseStep4}
\includegraphics[scale=1]{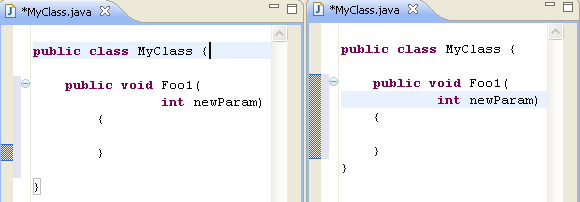}
\end{figure*}

\end{enumerate}

Although technically possible, collaboratively coding a tight piece of code,
like a method definition for instance, is usually not recommenced.

It is worth noting that in a typical SCM system, such a conflict would only be
discovered post factum, when a developer would try to commit an already
conflicting code version. The SCM system would then alert that an update should
be performed before a commit can take place. It is this update that will make
the developer aware of the conflict, and will force him to perform a manual
merge. In a CRTC system the conflict is discovered \textbf{before} it actually
intrudes the system, while in current SCM systems it is only discovered
\textbf{after} it is in the system.

The key concept of a CRTC system is that it is aware of all changes currently
carried by all developers. Thus, once the (chronologically) first developer
begins changing the method's name, the CRTC system locks the method element for
editing by other developers. When another developer tries to change the same
method, the CRTC system will notify him that the element he's trying to edit is
already being changed by another developer. He may then choose to undo his
changes (by using undo operations) and wait until the undergoing change is
complete before introducing his own. The CRTC system is able to serialize what
used to be concurrent, unsynchronized changes performed on the same elements, so
that they are no longer concurrent, and are in fact conflict free. However, the
serialization is so fine grained that we expect it to be practically transparent
to the developer, and he will only be aware of it in case it intervenes to
prevent a definite conflict.

The principles from the use case we've described apply to a wide variety of
changes: introducing new methods, changing existing method's name, changing
existing method's parameters, changing existing method's body, removing an
existing method introducing a new member variable, changing existing member's
name, removing an existing member. In general, a CRTC should support all code
editing operations available in a standard IDE.

\subsection{The code-as-text approach}

We next outline some of the common challenges that emerge in
collaborative editing in general. Issues arising in collaborative
text editing, along with a prototype system resolving them were
suggested by \cite{ConcurrencyControlInGroupwareSystems}.

Collaborative text editors share text documents. While seemingly
trivial, this turns out to be an important observation. Plain text
lies in the basis of both data and presentation layers. End users
see text characters, and perform editing operations on these
characters, which are then propagated to other users. In
collaborative code editors, one might assume a similar principle
where the text just happens to be code. However, this approach
raises a few issues.

In case the
characters being typed fail to form a valid statement (or any other legal
language element for that matter), rendering some local file state
unbuildable. Propagating these insertions would render the state of all users
unbuildable. In fact, the naive model of propagating code on a character by
character basis essentially means that whenever so much as one user enters a
buildbreaking state - all users are forced into a buildbreaking state as well.
Users who had nothing to do with the buildbreaking change per se will suffer the
consequences just as much. This seems unacceptable as it significantly harms
individual effectiveness and progress, failing to provide a basic level of
isolation for individual users. The desired behavior is to allow users to work
as independently as possible, making them less susceptible to intermediate build
breaking states brought about by others, while at the same time keeping them up
to date with recent code changes.

Suppose a user renames a method, but his code is not yet buildable, and is thus
not propagated. If in the meantime, another user happens to produce some new
code using the old method's name (since he is unaware of the change), a conflict
will arise. Once both users locally reach a buildable state and their changes
get propagated to each other, they will end up having two conflicting code
versions. It may be noted that this issue cannot be resolved on a text
characters level. The problem lies in the semantics of a given programming
language and can not be detected on the plain text level, a semantic awareness
must be present. In order to address semantic conflicts, the act of change
propagation must be aware of the syntax and semantics.

Syntax may also pose a challenge. Suppose a user writes a {\it for}
loop block. We have very little information on the corresponding AST
(Abstract Syntax Tree) node until the characters being typed sum up
to a legal {\it for} loop block that can be parsed. This
intermediate state, until the user is done typing, can not be
immediately propagated to other users. Doing so might (and most
probably will) render the file unbuildable (the unfinished {\it for}
loop will not compile). During an intermediate state, the mapping
between AST nodes and their textual representation may be
temporarily out of date, leaving the {\it for}'s inner block
unmapped. This may prevent the system from enforcing semantic rules
(such as "locking") on the {\it for} body statements until the whole
{\it for} is complete.

It may be observed that at times the presentation and underlying data model
states differ. These differences are usually the result of unbuildable code,
where the textual representation contains information not yet expressed by
existing AST nodes (and the semantic model it represents). It is this delta that
adds challenges to the propagation methods, and prevent from the naive,
immediate character by character propagation method to be truly effective and
practical.
Ideally, if the presentation model (i.e. the way code is presented to the
developer) and underlying model (i.e. the way code is manipulated by the
compiler) were the same, such as in the case of traditional plain text editing,
designing the propagation method would be simplified significantly. In such
cases the presentation data could be propagated in its raw form. This however,
may not turn out to be always feasible when dealing with code (as will be
discussed later on). A collaborative real time coding system should aim at
reducing these deltas, bringing the presentation and underlying data models
closer. Such a behavior might give users the illusion that code states move from
one buildable state to the next, as if the aforementioned deltas never existed.
The very essence of this idea is : the user should not have to know that there
is someone else sharing a file, unless they are both trying to modify the same
part, or unless close collaboration is desired \cite{EnsembleImplicitLocking}.

\subsection{Central vs. Distributed}

Similar to the pioneer system Jupiter
\cite{JupiterCollaborationSystem}, and its modern adaptation Google
Wave \cite{GoogleWave}, we base our model on a centralized entity
serving as a relay point through which all users communicate. This
greatly simplifies the setting, as many issues are reduced to the
case of two users, abstracting the case of $n$ users. Two users who
seemingly communicate with one another are actually communicating
with the central server, which relays their messages. At any given
time, only two parties are engaged in a concurrency synchronization
process, the central server and a client. Once the server is updated
with a change from a client, it propagates this change to the rest
of the clients. A real tool will need to insure that the main server
is not a single point of failure. This can be achieved using known
techniques, but it is out of the scope of this paper.

\subsection{Locking schemes}
\label{lockingSchemes}

In our work we concentrate on a solution based on a locking scheme. Locking
schemes may be classified into two main categories: optimistic and pessimistic.

\begin{itemize}
\item \textbf{Pessimistic locking} takes the view that users are highly likely to
corrupt each other's data, and that the only safe option is to serialize data
access, so at most one user has control of any piece of data at one time. This
ensures data integrity, but can severely reduce the amount of concurrent
activity the system can support  \cite{lockingSchemes}.
\item \textbf{Optimistic locking} takes the view
that such data collisions will occur rarely, so it's more important to allow
concurrent access than to lock out concurrent updates. The catch is that we
can't allow users to corrupt each other's data, so we have a problem if
concurrent updates are attempted. We must be able to detect competing updates,
and cause some updates to fail to preserve data integrity \cite{lockingSchemes}.
\end{itemize}

Optimistic locking schemes are considered better suited to
environments where communication latency is high but conflicts are
rare \cite{OptimalLockingIntegratedWithOperationalTransformation,
ConcurrencyControlEffectOnInterface}. Unlike most collaborative text
editing algorithms, which usually opt for an optimistic model with
no locking scheme, we believe that the task of collaborative real
time coding calls for a pessimistic model. However, we believe it is
dangerous, and therefore highly undesirable for developers to make
decisions based on stale code states, brought about as a by product
of the pessimistic model. In text editing this phenomenon is rather
common, and is resolved by the OT framework. Our fundamental
assumption is that while coding, a developer would rather wait
(obviously, within reason), than engage in a manual merge process
incurred by possible version conflicts caused by stale code states.
Moreover, conflicting operations are expected to occur rather
seldom, since although technically possible, collaboratively coding
a tight piece of code is usually not recommenced. In general,
conflicting versions is a well known issue in many source control
systems, which to some extent may be considered as a sort of
collaborative (non real time) coding environment. Conflicts are
usually resolved by preforming a merge. A merge process may be
either automatic or manual. If no conflicts are detected, an
automatic merge may take place, requiring no user intervention. In
case conflicts do arise, the system resorts to a manual merge,
conducted by the engaging the user.

Our efforts are therefore proactive, directed at
preventing conflicts in the first place. A trivial proactive solution would be
to allow only one developer to work on any given code file at a time. This is
however, a very coarsely granulated approach that greatly damages the real time
collaborative aspect of the system. In addition, file locking does not take
element dependency into account. Changing an element that is referenced in
another file without updating the reference will still cause a conflict. We seek
to introduce a consensus that will allow users to be up to date with recent
changes, while providing them with a real time collaborative environment and
a granularity that allows for tight cooperation.

It may be noted that the methods and models suggested so far have been
programming language agnostic as they did not rely on any language specific
attribute. The locking schemes, however, may have to be intimately
familiar with language specific attributes such as grammar and semantics. In the
suggested locking scheme, we establish the notion of element dependency.
Elements $E_1$, $E_2$ are {\em dependent} if one of the following holds:

\begin{enumerate}
 \item $E_1 = E_2$.
 \item $Parent(E_1) = Parent(E_2)$ in the AST.
\item $E_1$ is referenced by $E_2$ or vice versa.
\end{enumerate}

Our observation is that dependent elements (i.e., AST nodes) may not be subject to
concurrent editing.

The first case implies that no single element may be
concurrently edited.

The second case deals with concurrent editing of elements having a
common direct parent. The child elements commutativity property of a
common parent may depend on its type. If it is a class element for
example, its direct children can be reordered with no restriction
while preserving the semantics. If it is a method element on the
other hand, in the general case, its children (i.e., statement
elements) cannot be freely reordered as the overall method's
behavior may, and probably will, change. If semantic preservation
cannot be guaranteed in the general case, we enforce serialization,
taking the concurrency factor out of the equation. We disallow two
users to perform concurrent operations on dependent elements.
Instead, only one change maker is permitted to go through at a time.
One can think of various options as to the behavior a collaborative
system may adopt when a user encounters a situation where he's
denied immediate execution of his change. For instance, a trivial
one is to make him wait until his change may be executed (while
properly informing him of the circumstances). In principle, two
elements having a common parent whose children are semantically
commutative may be edited concurrently with the aid of OT.

We demonstrate the third case with a use case. Given a buildable state, and an
AST node $N$, we define $N$'s \emph{breakable set}, as all nodes that reference
$N$'s binding. One may think of the dependent elements as the set of elements
"used", or referenced, by a given element. Deducing the referenced elements
involves analyzing the element at hand according the language grammar. Moreover,
this analysis may be required on the fly, as users write code. It is therefore
crucial to determine the dependent elements as soon as possible in order for the
system to enforce the locking scheme in real time, while the user types in his
code. Any delay in doing so may result in a conflict, since as long as the
dependent elements remain unlocked, other users may change them concurrently
(which as mentioned, may result in a conflict). Let file1 include the definition
of a method named \emph{Foo}. Suppose developer1 changes the name of the
\emph{Foo} method to \emph{Foo1}, but his change has not been propagated to the
rest of the team, and developer2 adds a new method named \emph{UsingFoo} (even
in a different file, file2), which uses the old name, \emph{Foo} (developer2 is
unaware of the fact \emph{Foo1}s name has been changed). If developer1
propagates his change before developer2, the state will be rendered unbuildable
since the new code produced by developer2 will be invalid, due to the fact it
uses a method named \emph{Foo}, which no longer exists. It may be argued that in
this particular case, it would be better to allow developer2 to propagate
his changes first, making developer1 aware that he should also rename the new
usage of \emph{Foo} to \emph{Foo1}, but such a patch is merely a workaround, as
it fails to address the root cause of this problem.

We argue that the system is better off preventing this race
condition altogether, rather than resolving it. Locking also
provides the answer to less than trivial cases, such as the changing
a method's definition, which involves cascading changes, i.e.,
updating all callers accordingly. When such a change is detected by
the CRTC system, it should lock the method's element and its
dependent elements, which by definition include the callers.

Generally speaking, elements should retain their defining attributes (name,
type, scope, return type, parameter names, parameter types, parameter number,
etc) across the system while being edited. In other words, developers should not
be allowed (or should be notified at the very least) to concurrently manipulate,
use, or change, elements that are being edited. In our case, if \emph{Foo}'s
method definition was locked, developer2 would have been notified and made aware
of the problematic situation as soon as he tried to use \emph{Foo}'s old name,
and may have waited for develper1 to be done with his edit before going through
with his own. The decision whether to refrain from editing locked elements until
they are unlocked, or go through with the change (running the risk of
introducing conflicting versions) may be both left up to the user, or hard coded
in the system.

\subsection{User isolation level}

Although the ultimate goal of collaborative software is to give
users a variety of collaboration abilities, when coding, one's right for
autonomy should be taken into account as well . We believe it is no
coincidence that in many source control environments, the action of
sending one's code to the main repository and thus making it
publicly available, is called commit. Informally, when developers
commit their code they are expected to commit to its quality. It is
therefore common for developers to first perform unit tests on their
local workstations before committing code. Local testing reduces the
chance of bugs finding their way into public repositories and
eventually to the release version. In the collaborative coding model
described, users' operations are reflected in the common version
(i.e., the version everyone owns) as soon as the state is buildable.
It may be good practice to allow users some degree of isolation,
before propagating their changes to others. A user may choose to go
"off the record" whenever he wishes to delay the propagation of his
changes, despite the fact that technically they can be propagated
immediately. Local unit testing is great motivation for going off
record. However, going off record comes at a price. While off the
record, the shared code version (owned by the users who are on
record) evolves independently of the version owned by the user being
off record. This greatly increases the chance of introducing a
conflict once going back on record, since during the off record
period no restrictions (such as locking) are imposed on the changes
performed. We can clearly observe the tradeoffs between providing
users with close collaboration abilities and providing them with a
level of isolation. This comes as no surprise, as isolation and
collaboration lie at two opposite ends of the spectrum.

\section{A Collaborative Real Time Coding prototype\label{prototype}}


We demonstrate our approach for collaborative real time coding system with a
prototype, implemented as an Eclipse IDE plug in. Eclipse is an extensive open
source IDE, allowing developers to build their own plug in applications while
taking advantage of the vast Eclipse framework. In particular, Eclipse offers
the "Java Model", a set of classes that model the objects associated with
creating, editing, and building a Java program. The Java model classes are
defined in org.eclipse.jdt.core. These classes implement Java specific behavior
for resources and further decompose Java resources into model elements. The Java
development tools (JDT) uses an in-memory object model to represent the
structure of a Java program. This structure is derived from the project's class
path. The model is hierarchical, elements of a program can be decomposed into
child elements \cite{JavaModel}.

Our CRTC plug in, uses the Java Model offered by Eclipse in order to
be notified of changes made to the model representing the program
structure. These changes may include various operations, such as
introducing, deleting and changing Java elements like classes,
methods, member variables and more. The Java Model plays an
important role in tracking changes on a semantical level, rather
than observing textual changes. For instance, the Java Model enables
us to be notified of a new method being introduced, rather than of a
stream of characters representing this method's code being typed
into the IDE. Once the collaborative real time coding system is made
aware of semantic changes, it's able to propagate these changes to
all clients while retaining their semantic meaning, as opposed to
plain text propagation that is common to the code-as-test approach.
Our plug in also uses the IProblemRequestor interface, a callback
interface for receiving Java problems as they are discovered by some
Java operation \cite{IProblemRequestor}. Using this callback, we're
able to integrate with the error detection framework of Eclipse,
which is able to report errors in real time, on the fly, while the
resource (in our case, the Java file) at hand is being modified,
before it has even been saved. The real time error detection ability
is crucial to a CRTC system as it is tightly linked with code
propagation between clients. As previously noted, a CRTC system
strives to refrain from propagating code changes as long as the file
is unbuildable. It is therefore important to detect unbuildable
states as soon as possible. We demonstrate this idea in our
prototype, which does not wait for the code file to be saved in
order to process and propagate code. This may be witnessed by the
asterisk symbol near the file name at the top of the editing tab in
the Eclipse IDE. The asterisk symbol indicates that the file at hand
has not been saved yet and all changes are currently in memory
buffer; see figures \ref{fig:UseCaseStep1}, \ref{fig:UseCaseStep2},
\ref{fig:UseCaseStep3} and \ref{fig:UseCaseStep4}.

We conducted our testing and experimenting in a setup consisting of virtual
machines (VM), created by the Oracle VM VirtualBox \cite{OracleVM} application.
In this setup we had a server machine and two client machines, simulating two
developers working on a common codebase in the Eclipse IDE.

Our CRTC prototype supports the following scenarios:
\begin{itemize}
\item Introducing new methods
\item Changing existing method's name
\item Changing existing method's parameters
\item Changing existing method's body
\item Removing an existing method
\item Introducing a new member variable
\item Changing existing member's name
\item Removing an existing member
\end{itemize}

Although the prototype described demonstrates some key aspects of CRTC, few
issues remain to be further considered. For instance, while we prevent
syntactically invalid state propagation, we have not yet experimented with
semantically invalid states in our prototype. If a particular code change is
composed of a sequence of smaller changes, where each intermediate change in
syntactically valid, propagation may take place individually after each
individual change. However, it may be the case, that the intermediate changes
are semantically invalid, and do not represent the meaning of the intended
change. The current propagation method does not take such a scenario into
account. A more sophisticated propagation method might need to be explored,
for instance, one that propagates AST nodes instead of textual elements.

\section{Conclusions and Future Work\label{conc}}
\subsection{Discussion}

In this work we introduce the term Collaborative Real Time Coding
(CRTC), that describes the concept of real time collaboration and
code sharing between multiple programmers working on the same
software project. We've described the principles we believe a CRTC
system should follow in order to ensure conflict free collaboration
and provide near real time code propagation to all parties. We also
suggested a proof of concept by means of implementing a prototype
CRTC plug in for the Eclipse IDE. This prototype captured the
essence of a CRTC system and demonstrated how real time coding may
take place using a modern IDE.

The approach we propose, and the way the prototype works is a
radical departure from the common way software engineers work. Our
solution may be dismissed by some people as too radical, but we view
this as a first step in exploring possible options for people to
cooperate. We do believe that better interaction between individual
engineers working on a common project is needed, and that our
approach offers a basis for such interaction. In Hegelian terms, if
the conventional way of working with an SCM is the thesis, then our
proposal may be considered as an antithesis. Further work is
required to fully understand the implications of our approach, and
to see how best to make use of them. Hopefully, a synthesis will
emerge, which combines the two approaches.

\subsection{Future Work}

We believe it is worth while to further explore the possibility of
redesigning standard SCM systems into CRTC systems. This includes
accounting for the common SCM features (version history, check-in
and check-out operations, main repository, etc.) in addition to
incorporating the new, real time capabilities, into a unified real
time collaboration SCM tool, or even a real time collaboration IDE
with a built in SCM support.

It also seems beneficial to explore how CRTC can enhance the overall
collaboration in a software development team, for instance by incorporating
group awareness tools \cite{ERV} and providing developers with the ability to be
more attentive to the work being done by their colleagues.

Since CRTC operates in a fine grained manner, it presents the
opportunity to implement a variety of features operating on an
element level basis. For instance, if a developer wants to avoid
editing an element in case it is locked, it may be useful to allow
him to register for notifications on changes pertaining to that
particular element. He might, for instance, want to be notified as
soon as the given element becomes unlocked so he can perform the
change we was intending. It may be also helpful should a developer
serve as a gatekeeper and require to be notified whenever certain
elements are changed so he could inspect and review the changes.
This in turn may lead to an access control mechanisms enforced on
certain elements and/or developers. One can have the option to limit
access (be it read or write privileges) to certain elements and/or
developers.

CRTC opens many doors to future research in terms of supporting environment,
coding conventions, work procedures and IDE capabilities. We believe that in
light of CRTC's novel approach to collaboration between software developers,
certain approaches may need to be extended in order to fully utilize the
benefits of CRTC.

Software methodologies are also of great effect on CRTC. We believe it is
important to further research and gain experience as to where in the application
life cycle does CRTC fit best. Many directions remain to be further explored,
such as for instance , how does CRTC fit in modern methodologies like Agile? Or
the more mature ones like Waterfall?
CRCT may also have a significant impact on
distributed software development, allowing methods like extreme programming to
take place in a geographically separate locations.

Disciplines like Refactoring for instance, may potentially have great interactions with CRTC.
Since refactoring is essentially a set of changes introduced to a given code,
it's worth exploring how it affects CRTC capabilities. It may be the case, that
using certain refactorings instead of manually performing equivalent changes,
may aid a CRTC system to enforce correctness even when faced with the more
complex collaboration scenarios.

Another discipline that may potentially interact with CRTC is unit testing.
Since in a CRTC system there is a central server that's aware of all changes in
near real time, it may be used to run unit tests and verify no regression takes
place in a continuous manner. Unit tests may also be incorporated into the
propagation trigger, so that before propagating any local code it will be unit
tested automatically by the CRTC system.

Finally, it is clear that some user studies and experiments will be
needed to evaluate this approach and the various alternatives of its
use.

\bibliographystyle{IEEEtran}
\renewcommand\refname{Bibliography}
\bibliography{CRTC}

\end{document}